\documentclass[10pt]{article}
\usepackage{epsfig}
\usepackage{mathtools}
\usepackage{amsfonts}
\usepackage{amssymb}
\usepackage{amsmath}
\usepackage{listings}

\usepackage{color}
\usepackage{slashed}
\usepackage{cite}
\usepackage{caption}
\usepackage{subcaption}

\usepackage{mmacells}

\usepackage{geometry} 
\usepackage{graphicx}
\graphicspath{{pictures/}}

\usepackage{multirow}           
\usepackage{array}              
\def\ep  {\varepsilon}
\def\E4  {\mathcal{E}_4}

\def\pFq   {~_p F_q }

\newcommand{\J}{{\bf J}}
\newcommand{\M}{{\bf M}}
\newcommand{\A}{{\bf A}}
\newcommand{\B}{{\bf B}}
\newcommand{\C}{{\bf C}}

\newcommand{\Mt}{{\tilde{\bf M}}}
\newcommand{\Mb}{{\bar{\bf M}}}
\newcommand{\Jt}{{\tilde{\bf J}}}
\newcommand{\T}{{\bf T}}
\newcommand{\Ladapter}{{\bf L}}
\newcommand{\uvec}{{\bf u}}
\newcommand{\wvec}{{\bf w}}

\newcommand{\OO}{\mathcal{O}}

\begin{document}

\begin{center}

\vspace{3cm}

{\bf \Large Expansion of hypergeometric functions in terms of polylogarithms with nontrivial variable change} \vspace{1cm}

{\large M.A. Bezuglov$^{1,2}$ and A.I. Onishchenko$^{1,2,3}$}\vspace{0.5cm}

{\it $^1$Bogoliubov Laboratory of Theoretical Physics, Joint
	Institute for Nuclear Research, Dubna, Russia, \\
	$^2$ Budker Institute of Nuclear Physics, Novosibirsk, Russia,\\
	$^3$Skobeltsyn Institute of Nuclear Physics, Moscow State University, Moscow, Russia}\vspace{1cm}
\end{center}

\begin{abstract}
Hypergeometric functions of one and many variables play an important role in various branches of modern physics and mathematics. Often we have hypergeometric functions with indices linear dependent on a small parameter with respect to which one needs to perform Laurent expansions. Moreover such expansions are desirable to be expressed in terms of well known functions which can be evaluated with arbitrary precision. To solve this problem we use the differential equation method and the reduction of corresponding differential systems to canonical basis. Specifically we will be interested in the generalized hypergeometric functions of one variable together with Appell and Lauricella functions and their expansions in terms of Goncharov  polylogarithms. Particular attention will be given to the case of rational indices of considered hypergeometric functions when the reduction to canonical basis involves nontrivial variable change.
The article comes with a Mathematica package \texttt{Diogenes}, which provides algorithmic implementation of the required steps.  
\end{abstract}

\begin{center}
Keywords: Generalized hypergeometric functions, Appell and Lauricella functions.
\end{center}

\newpage

\tableofcontents{}\vspace{0.5cm}

\renewcommand{\theequation}{\thesection.\arabic{equation}}

\section{Introduction}

Hypergeometric functions of one and many variables play an important role in 
many different areas of both physics and mathematics. In particular they frequently appear in the calculation of multi-loop Feynman diagrams in quantum field theory. There are various methods for obtaining solutions for scalar Feynman integrals in terms of hypergeometric functions of one and many variables. Among such techniques are Mellin-Barns method\footnote{See \cite{Weinzierl:2022eaz,Dubovyk:2022obc,smirnov2006feynman} for introduction and references to original works. There is also a bunch of packages to work with Mellin-Barnes integrals \cite{Belitsky:2022gba,Ananthanarayan:2020fhl,Ochman:2015fho,Smirnov:2009up,Gluza:2007rt,Czakon:2005rk}.
}, the DRA method \cite{Tarasov:2006nk,  Lee:2009dh, Lee:2012hp}, method of functional equations \cite{Tarasov:2022clb} and the exact Frobenius method \cite{Bezuglov:2022npo, Bezuglov:2021tax, Blumlein:2021hbq}. One should also mention the connection of Feynman integrals to GKZ systems\footnote{See \cite{Matsubara-Heo:2023ylc,Vanhove:2018mto} for an introduction and general overview.}  \cite{Gelfand1992GeneralHS,Gelfand1994DiscriminantsRA,Gelfand:1990bua,beukers2013monodromy,Kalmykov:2012rr,delaCruz:2019skx,Klausen:2019hrg,Ananthanarayan:2022ntm}. In all these cases one obtains hypergeometric functions with indices linear dependent on the parameter of dimensional regularization $\ep$. In practical applications we are further required to expand these functions in Laurent series at small values of $\ep$ up to some specified order. Such expansions were already extensively studied in literature, see for example \cite{Davydychev:2003mv,Kalmykov:2004kg,Kalmykov:2006pu,Kalmykov:2007pf,Kalmykov:2008ge,Greynat:2013hox,Greynat:2013zqa,Greynat:2014jsa,Moch:2001zr,Weinzierl:2004bn,Yost:2011wk,Bytev:2012ud,Kalmykov:2020cqz,Bera:2022ecn}. Moreover, automatic packages for such expansions exist \cite{Huber:2005yg,Huber:2007dx,Moch:2005uc,Weinzierl:2002hv,Ablinger:2013cf,Huang:2012qz,Bera:2023pyz}. Nevertheless, there are a lot of open questions left. For example, when a given hypergeometric function can be expanded in terms of polylogarithms or some other functions and do we understand a systematic procedure in each case? 

We certainly would like to have expansions in terms of well defined functions, which can be finally evaluated at general parameter values with as good numerical precision as possible. The Goncharov multiple polylogarithms \cite{goncharov2,goncharov3} are certainly among such functions. Their properties are well studied \cite{goncharov4,goncharov5Symbool,Duhr1,Duhr2,duhr2012polygons,Vollinga:2004sn} and we have a number of Mathematica packages, such as HyperInt \cite{HyperInt}, MPL \cite{Bogner:2015nda} and PolyLogTools \cite{duhr2019polylogtools} to work with them. In addition, it is very important that they can be calculated numerically with arbitrary precision \cite{Vollinga:2004sn, Naterop:2019xaf}.

The goal of this paper is to apply methods originally developed for calculating multiloop Feynman integrals for $\ep$-expansion of hypergeometric functions of one and many variables, which indices are linear dependent on $\ep$. In particular, we will be using the method of differential equations \cite{KOTIKOV1991158,kotikov1991differential,kotikov1991differential2,remiddi1997differential,gehrmann2000differential, argeri2007feynman,henn2015lectures} and the
reduction of hypergeometric differential systems to $\ep$-form \cite{epform1,epform2,epform-criterium}. This possibility in the case of generalized hypergeometric functions was already mentioned in \cite{Bezuglov:2022npo} and extensively studied recently for the expansion of hypergeometric functions of both one and many variables in \cite{Bera:2023pyz}.
However, in \cite{Bera:2023pyz} only the case of integer hypergeometric indices was considered. Here, we will also consider rational indices for which an $\ep$-expansion in terms of multiple polylogarithms is still possible. The expansion procedure was realized in Mathematica package \texttt{Diogenes}. The latter allows to perform expansions in some cases not covered by the existing packages and we hope it will be useful for community. The package is self contained and does not require installation of additional software besides Wolfram Mathematica system itself.

The remainder of the paper is organized as follows. In the next section \ref{sec:pFqInG} we extensively study the $\ep$-expansion of generalized hypergeometric functions of one variable. Particular attention is paid to nontrivial variable change, which allows for polylogarithmic expansion of hypergeometric functions with rational indices. Sections \ref{sec:AppellInG} and \ref{sec:LauricellaInG} deal with the $\ep$-expansion of Appell and Lauricella functions. Here the study is less complete compared to the case of generalized hypergeometric functions. Still, we consider several cases with nontrivial variable change leading to resulting expansion in terms of multiple polylogarithms. The appendices are introduced to set up notation for multiple polylogarithms including cyclotomic, recall main steps of the reduction to $\ep$-form and provide examples of \texttt{Diogenes} package usage.  Finally, in section \ref{section::Сonclusions} we give comments on future research directions.

\section{Expansion of generalized hypergeometric functions}
\label{sec:pFqInG}

It is  natural to start the discussion of $\ep$-expansion of hypergeometric functions with generalized hypergeometric functions of one variable. The latter may be defined with the following series representation
\begin{equation}
\label{eq:pFqDef}
~_p F_q \left(
\begin{array}{c}
a_1,\dots,a_p  \\ b_1, \dots, b_q
\end{array} \Big|z\right)= \sum\limits_{n=0}^{\infty}\frac{(a_1)_n,\dots , (a_p)_n}{(b_1)_n,\dots , (b_q)_n}\frac{z^n}{n!}
\end{equation}
where $(~)_n$  denotes Pochhammer symbol. Here we will always consider the particularly interesting case when $p=q+1$, for which the above series converges in the region $|z| < 1$. If in addition $\text{Re}\left(\sum b_i - \sum a_i\right) > 0$ then the convergence region will also include $|z|=1$. By default, we will always assume that the indices  $a_i$ and $b_j$ are linearly dependent on a small parameter $\ep$. Also, sometimes, to shorten notation we will omit the function indices and simply write $_p F_q$. To obtain $\ep$-expansion of $_pF_q$ functions \eqref{eq:pFqDef} we start with the ordinary differential equation it obeys
\begin{equation}
\label{eq:DifferentialEquation}
\big[z(\theta+a_1)(\theta+a_2)\dots(\theta+a_p)-\theta(\theta+b_1-1)(\theta + b_2-1)\dots (\theta + b_q -1)\big]~_p F_q=0
\end{equation}
where $\theta = z \frac{d}{dz}$. This is linear homogeneous equation of $p$-th order, which gives us possibility to write down linear matrix differential system for a vector  of $_p F_q$ function together with its derivatives over $z$ up to $q$-th order. To do this we define a vector of $p$ functions 
\begin{equation}
\label{eq:BasisForPFQ}
\J=\{f_0,f_1,\dots,f_q\}
\end{equation}
where
\begin{equation}
f_0 = ~_p F_q, \qquad
f_n = \theta(\theta-1)\dots(\theta -n+1)f_0 = z^n \frac{d^n f_0}{dz^n}.
\end{equation}
and consider first order differential equations for its components. For $n < q$ the latter are naturally given by  
\begin{equation}
\frac{d}{dz}f_n = \frac{n}{z}f_n + \frac{1}{z}f_{n+1}
\label{eq:DiffEqForBasisElementsFn}
\end{equation}
and for $n = q$ the corresponding differential equation can be obtained directly from the original equation  \eqref{eq:DifferentialEquation}. As a result, we get a system of differential equations in the form
\begin{equation}
\frac{d\J}{dz} = \M\cdot\J\, ,\quad 
\M = \frac{\A}{z} + \frac{\B}{z-1}\, ,
\label{eq:DiffEqFuchsian}
\end{equation}
where $\M$, $\A$ and $\B$ are some $p \times p$ matrices. This differential system is Fuchsian and has three first order poles at $0$, $1$ and $\infty$. The boundary conditions for this system easily follow from the definition of $_pF_q$ function \eqref{eq:pFqDef} and we get  
\begin{equation}
\J \big|_{z = 0} = \left\{1, 0, \dots, 0\right\}.
\label{eq:BoundaryConditionspFq}
\end{equation}

To solve this system perturbatively in $\ep$ we use the reduction\footnote{See appendix \ref{appendix::DE-epform} for a brief exposition of relevant ideas.} to so called $\ep$-form \cite{epform1,epform2,epform-criterium}. That is we are using Lee algorithm \cite{epform2} to find a rational transformation matrix $\T$ to the canonical basis $\Jt = \T^{-1}\J$, such that $\ep$-dependence in the transformed differential system factorizes and we have
\begin{equation}
\frac{d\Jt}{dx} = \ep\Mt\cdot\Jt\, ,  \label{eq:pFq-epform}
\end{equation}
where $\Mt$ matrix does not already  depend on $\ep$. It should be noted that such transformation does not always exists. In particular it does not exist in the case when $\J$ is the section of non trivial vector bundle on Riemann sphere over $z$ \cite{epform-criterium}. Here, we will be interested in the cases when it does. Moreover, to find transformation matrix $\T$ we will allow for a rational change of variable $z$. Provided the transformation matrix is found and original differential system can be reduced to $\ep$-form the solution of the resulting differential system \eqref{eq:pFq-epform} is easy and takes the form
\begin{equation}
\J = \T\cdot \text{Pexp} \left[
\ep \int_0^z \Mt(z') dz'
\right]\cdot \Ladapter\cdot \{1,0,\ldots 0\}^{\top}\, \label{eq:pFq-polylog-solution}
\end{equation}
where the boundary conditions for $\Jt$ vector at $z=0$ are related to the boundary conditions of the original $\J$ vector with a rational in $z$
adapter matrix $\Ladapter$. Expanding the obtained solution in \eqref{eq:pFq-polylog-solution} in $\ep$ will then result in the expansion of the original $_pF_q$ function in terms of multiple polylogarithms\footnote{See appendix \ref{appendix::MPLs} for our notation for Goncharov multiple polylogarithms.}.

Let us now discuss in detail when the transformation of the original differential system to $\ep$-form is possible after all. The knowledge of how balance transformations in Lee algorithm change the differential system under consideration tells us that our ability to find required transformation matrix depends solely on the eigenvalues of matrices $\A$, $\B$ and $\C = -\A-\B$ evaluated at $\ep = 0$, where   $\C$ is the residue of $\M$  matrix at infinity. The latter depend on indices of our $_pF_q$ function and can be explicitly written as  
\begin{align}
\mathbf{A}&:~\left\{0,1-b_1,\dots, 1-b_q\right\},
\label{eq::eigen0}
\\
\mathbf{B}&:~\left\{\underbrace{0,\dots,0}_{q},-q-\sum\limits_{i=1}^{p}a_i+\sum\limits_{i=1}^{q}b_i\right\},
\label{eq::eigen1}
\\
\mathbf{C}&:~\left\{a_1,a_2,\dots,a_p\right\}.
\label{eq::eigenInf}
\end{align}
and the sum of all eigenvalues is equal to zero.

If all eigenvalues are integers, that is all $a$ and $b$ indices\footnote{Here and in subsequent discussion we refer to values of indices and matrix eigenvalues evaluated at $\ep = 0$.} are integers, then we can directly apply Lee algorithm to balance eigenvalues and reduce differential system to $\ep$-form. However, in general in applications the values of $_pF_q$ indices are rational. In this case the Lee algorithm does not immediately applies. The reason is that the balance transformations \eqref{eq:balance} employed by the algorithm can shift matrix eigenvalues only by plus or minus one. It is possible that this problem can be solved by a rational variable change\footnote{Of course, this transformation is rational only in one direction.} so that the eigenvalues of the differential system in new variable become integer. Empirically we have found that such variable change can be found in the cases shown in Table \ref{tab:polylog-cases}.
For practical calculations to accelerate reduction procedure in these cases it is convenient to use balance transformations both before and after the variable change. In the case E it was found that the variable change better to perform in two stages $z \rightarrow 1+ \tilde{z}^2$ and $\tilde{z} \rightarrow i(1+z_4^2)/(1-z_4^2)$ with additional eigenvalue balancing in between. It can also happen that the variable change in these cases does not necessary leads to $\ep$-form and to achieve the latter one needs to apply additional restrictions. Denoting $k$ as the number of non-integer upper indices $k = \#\{ a_i| i=1,\dots,q ; a_i \notin \mathbb{Z} \}$, $l$ as the number of non-integer lower indices $l = \#\{ b_j| j=1,\dots,p ; b_j \notin \mathbb{Z} \}$ and $n$  as the least common denominator for all non-integer indices we have empirically found that the reduction to the $\ep$-form by our algorithm is not possible when in addition one of the following conditions holds  
\begin{table}
\begin{center}
\begin{tabular}{||c c c c c||} 
\hline
Case & $\mathbf{A}$ eigenvalues & $\mathbf{B}$ eigenvalues & $\mathbf{C}$ eigenvalues & variable change \\ [0.5ex] 
\hline\hline
$A$ & $m_a+q_a\ep$ & $m_b+q_b\ep$ & $m_c+q_c\ep$ & none \\ 
\hline
$B$ & $\frac{k_a}{n} + m_a+q_a\ep$ & $\frac{k_b}{n} + m_b+q_b\ep$ & $m_c+q_c\ep$ & $z \rightarrow \frac{z_1^n}{1+z_1^n}$ \\
\hline
$C$ & $\frac{k_a}{n} + m_a+q_a\ep$ & $m_b+q_b\ep$ & $\frac{k_c}{n} +m_c+q_c\ep$ & $z \rightarrow z_2^n$ \\
\hline
$D$ & $ m_a+q_a\ep$ & $\frac{k_b}{n} +m_b+q_b\ep$ & $\frac{k_c}{n} +m_c+q_c\ep$ & $z \rightarrow 1 -z_3^n$ \\
\hline
$E$ &  $ \frac{1}{2}+m_a+q_a\ep$ & $\frac{1}{2}+m_b+q_b\ep$ & $\frac{1}{2}+m_c+q_c\ep$ & $z \rightarrow -\frac{4 z_4^2}{(z_4^2-1)^2}$ \\
\hline
$F^{*}$ &  $ \frac{k_a}{n}+m_a+q_a\ep$ & $\frac{k_a}{n}+m_b+q_b\ep$ & $\frac{k_a}{n}+m_c+q_c\ep$ & $z \rightarrow \frac{z_1^n}{1+z_1^n}$ \\ [1ex] 
\hline
\end{tabular}
\end{center}
\caption{The cases when differential system for $_pF_q$ function can be reduced to $\ep$-form. Here $m_i, k_i\in \mathbb{Z}$, $n\in \mathbb{N}$ and $|k_i| < n$. See additional clarifications about $F^{*}$ case in the main text.}
\label{tab:polylog-cases}
\end{table}

\begin{enumerate}
	\item $n = 2$ and $|k-l|\ge 2$.
	\item $n > 2$ and ($k \ge 2$ or $l \ge 2$) and $\{a_i-a_1, b_j-a_1| i=1,\dots,q, j=1,\dots,p \} \notin \mathbb{Z} $.
	\item $n > 2$, $k = l = 1$ and $a_i - b_j \notin \mathbb{Z} $ where $a_i$ and $b_j$ are two non-integer indices.
\end{enumerate}  

Case F requires additional explanation. In cases B-E variable transformations contained in Table \ref{tab:polylog-cases} automatically convert all eigenvalues to integers. However, this does not work for the case F.  Moreover, in this case such rational transformation does not exist at all. Consider the variable change $z = p(z')/q(z')$ where $p$ and $z$ are coprime polynomials. If the differential system has rational eigenvalues with least common denominator at singular points $z_i$ equal to $n$ then this variable change which will convert all eigenvalues to integers should satisfy the following restrictions on $p$ and $q$ polynomials \cite{epform-criterium}:
\begin{equation}
\beta_i p(z')-\alpha_i q(z') = \tilde{p}_i^{n}(z')
\end{equation}
where $\tilde{p}_i(z')$ are some other polynomials and $[\alpha_i:\beta_i]$ are the homogeneous coordinates of $z_i = \alpha_i/\beta_i$ points. In the case of our three singular points 
$z_1 = 0,~ z_2 = 1$ and $z_3 = \infty$ we thus get
\begin{align}
p(z') & = \tilde{p}_1^{n}(z')\\
p(z') - q(z') & = \tilde{p}_2^{n}(z')\\
q(z') & = \tilde{p}_3^{n}(z')
\end{align}
and as consequence 
\begin{equation}
\label{eq:LastFermatForPolinomials}
\tilde{p}_2^{n}(z')+\tilde{p}_3^{n}(z')=\tilde{p}_1^{n}(z')
\end{equation}
In the case $n = 2$ this equation has a solution of $\tilde{p}_1(z') = (1+(z')^2), ~ \tilde{p}_2(z') = (1-(z')^2)$ and $\tilde{p}_3(z') = 2z'$ corresponding to case E. But for $n >2$, as we know from Fermat's theorem for polynomials, this equation does not have non-zero polynomial solutions. Thus, there is no rational variable transformation that makes all eigenvalues at three singular points integer if $n >2$. However, no one forbids us to use non-rational transformations, namely the non-rational transformation matrix $T$. Consider for example transformation 
$
T = (z - z_i)^{a} \mathcal{I}\, 
$,  where $ \mathcal{I}$ is identity matrix. From \eqref{eq:DEGeneral2} two important properties of this transformation can be noted. First, such transformation leaves the $\M$ matrix of the differential system rational and, secondly, it shifts all  eigenvalues at singular point $z_i$ by $a$. Thus, if at one singular point all eigenvalues have the same non-integer part, then it is possible using this transformation to get rid of it at once. In the case of generalized hypergeometric function at hand this is only possible if all its upper indices $a_i$ have the same non-integer part, that is $\{a_i-a_1| i=1,\dots,q \} \in \mathbb{Z} $. If this condition is met, then using transformation matrix $T = z^{-\text{frac}(a_1)}\mathcal{I}$ we can make the eigenvalues integer at the point $z = \infty$ and then apply the transformation $z \rightarrow z_1^n/(1+z_1^n)$ to make the eigenvalues integer at points $z = 0,1$.  It is important to note that the resulting solution will have the form $y(z_1)*\text{MPLs}(z_1)$ where $y(z_1)$ is some (hyper)elliptic curve. Thus, although the (hyper)elliptic curve itself is not included in the iterated integrals as a kernel, it still remains as a factor. 

The reduction of differential system for generalized hypergeometric functions to $\ep$-form involves balancing its eigenvalues at three singular points. The latter are linear dependent on hypergeometric function indices, which can be also shifted with the use of  differential shift operators $\mathcal{DF}_{a}^z = \frac{z}{a}\frac{d}{dz}+1$ $(a \ne 0)$ as
\begin{align}
\mathcal{DF}_{a_j}^z ~_p F_q \left(
\begin{matrix}
a_1,\dots, a_j,\dots,a_p  \\ b_1, \dots, b_q 
\end{matrix} \Bigg| z\right) &=
~_p F_q \left(
\begin{matrix}
a_1,\dots, a_j+1,\dots,a_p  \\ b_1, \dots, b_q 
\end{matrix} \Bigg| z\right)
\label{eq:DifferentialRelationsA}
\\
\mathcal{DF}_{b_k-1}^z ~_p F_q \left(
\begin{matrix}
a_1,\dots ,a_p  \\ b_1, \dots, b_k, \dots, b_q
\end{matrix}
\Bigg| z\right) &=
~_p F_q \left(\begin{matrix}
a_1,\dots ,a_p  \\ b_1, \dots, b_k-1, \dots, b_q
\end{matrix} \Bigg| z\right)
\label{eq:DifferentialRelationsB}
\end{align}
such that
\begin{equation}
~_p F_q \left(
\begin{array}{c}
a_1,\dots,a_p  \\ b_1, \dots, b_q
\end{array} \Big|z\right)
=\mathcal{DF}_{s_1}^x\dots \mathcal{DF}_{s_l}^x
~_p F_q \left(
\begin{matrix}
\tilde{a}_1,\dots,\tilde{a}_p  \\ \tilde{b}_1, \dots, \tilde{b}_q
\end{matrix}
\Bigg| z\right)\, , \label{eq:DifferentialRelationFinal}
\end{equation}
where $\tilde{a}_i$ and $\tilde{b}_j$ is a new set of indices. So, 
before using balancing procedure one may shift indices of hypergeometric function to smaller values, obtain its expansion in terms of MPLs as a result of the described procedure and at the end use the relation \eqref{eq:DifferentialRelationFinal} to obtain expansion for the original $_pF_q$ function. In some cases this shift may increase performance. Such differential shift operators were also used in \cite{Bera:2023pyz} using the HYPERDIRE package \cite{HYPERDIRE1,HYPERDIRE2,HYPERDIRE3,HYPERDIRE4}

\subsection{Example for  case C}

Now let us illustrate the above expansion procedure on some particular examples. First consider the $\ep$-expansion of
\begin{equation}
~_3F_2 \left(\begin{matrix}
\frac{1}{2}, 1, -1+2\ep \\ 2-\ep, \frac{1}{2}+\ep
\end{matrix}\Bigg| z \right)\, .
\end{equation}
The differential system in this case is given by
\begin{equation}
\frac{d\J}{dz} = \left(
\frac{\A}{z} + \frac{\B}{z-1}
\right)\J\, , \label{eq:diff-system-example}
\end{equation}
where
\begin{equation}
\A = \begin{pmatrix}
0 & 1 & 0 \\0 & 1 & 1 \\
0 & \frac{2\ep^2-3\ep-2}{2} & -\frac{3}{2}
\end{pmatrix}\, , \quad \B = \begin{pmatrix}
0 & 0 & 0 \\
0 & 0 & 0 \\
\frac{1-2\ep}{2} & \frac{1-7\ep-2\ep^2}{2} & -2\ep
\end{pmatrix}\, 
\end{equation}
and $\C = -\A -\B$. The eigenvalues of these matrices are
\begin{equation}
\A : \left\{
0, -1+\ep, \frac{1-2\ep}{2}
\right\}\, , \quad \B : \left\{
0, 0, -2\ep
\right\}\, , \quad \C = \left\{
1, \frac{1}{2}, -1+2\ep
\right\}
\end{equation}
and thus according to our classification we are dealing with case C. Making variable change $z\to z_2^2$ and constructing required balance transformations the differential system \eqref{eq:diff-system-example} can be reduced to the following $\ep$-form ($\Jt = \T\cdot\J$):
\begin{equation}
\frac{d\Jt}{dz_2} = \ep \left(
\frac{\Mt_0}{z_2} + \frac{\Mt_1}{z_2-1} + \frac{\Mt_{-1}}{z_2+1}
\right)\cdot\Jt\, ,
\end{equation}
where
\begin{equation}
\Mt_0 = \begin{pmatrix}
2 & 0 & 0 \\0 & 0 & 0 \\
0 & 0 & -2
\end{pmatrix} ,\quad \Mt_1 = \begin{pmatrix}
-3 & 1 & -\frac{1}{2} \\
0 & 0 & \\
6 & -2 & 1
\end{pmatrix} ,\quad \Mt_{-1} = \begin{pmatrix}
-3 & 1 & -\frac{1}{2} \\
0 & 0 & 0 \\
-6 & 2 & 1
\end{pmatrix} .
\end{equation}
Now using the expressions for the first row of transformation matrix $\T$
\begin{equation}
\T_{(1,*)} = \left\{
\frac{1+z^4 (1-4\ep)-4\ep+z^2(-2+4\ep)}{z^2 (\ep-1)}, \frac{2-5\ep+z^2(4\ep-1)}{2(\ep-1)}, \frac{(1+z^2)\ep}{z(1-\ep)}
\right\}
\end{equation}
and adapter matrix $\Ladapter$
\begin{equation}
\Ladapter = \begin{pmatrix}
0 & 0 & \frac{\ep-1}{1-4\ep} \\
\frac{2 (1-\ep)^2}{(2-3\ep)(3\ep-1)} & 0 & 0 \\
0 & \frac{(1-\ep)(3-4\ep)(1-4\ep)}{4\ep (2-3\ep)(1-3\ep)} & 0
\end{pmatrix}
\end{equation}
we get from \eqref{eq:pFq-polylog-solution} an expansion of our hypergeometric function
\begin{equation}
~_3F_2 \left(\begin{matrix}
\frac{1}{2}, 1, -1+2\ep \\ 2-\ep, \frac{1}{2}+\ep
\end{matrix}\Bigg| z \right) = 1 - \frac{z}{2} + \ep\left\{
1+\frac{z}{4} + \frac{(1-z)^2}{z} (G_{-1}(\sqrt{z}) + G_1(\sqrt{z})) 
\right\} + \OO (\ep^2)
\end{equation}

\subsection{Example for case E}
Next. consider the $\ep$-expansion of 
\begin{equation}
~_3 F_2 \left(\begin{array}{c} 1,\frac{\varepsilon +1}{2},\frac{\varepsilon}{2}  \\ \frac{1-\varepsilon }{2},\frac{\varepsilon +3}{2}  \end{array} \Big|z\right), 
\end{equation}
which appeared n the problem of calculating two-loop corrections to the parapositronium decay \cite{Bezuglov:2022npo}. The differential system in this case is given by
\begin{equation}
\frac{d}{dz}\J = \left(\frac{\mathbf{A}}{z}+\frac{\mathbf{B}}{z-1}\right)\J
\end{equation}
where
\begin{equation}
\mathbf{A} = \left(
\begin{array}{ccc}
0 & 1 & 0 \\
0 & 1 & 1 \\
0 & \frac{1}{4} \left(\varepsilon ^2+2 \varepsilon -3\right) & -1 \\
\end{array}
\right), \qquad 
\mathbf{B} =\left(
\begin{array}{ccc}
0 & 0 & 0 \\
0 & 0 & 0 \\
-\frac{1}{4} \varepsilon  (\varepsilon +1) & \frac{1}{4} \left(-2
\varepsilon ^2-11 \varepsilon -9\right) & -\varepsilon
-\frac{3}{2} \\
\end{array}
\right)
\end{equation}
and $\mathbf{C} = -\mathbf{A}-\mathbf{B}$. The vector of boundary conditions is obviously
\begin{equation}
\J \bigg|_{z = 0} = \left\{1,0,0\right\}.
\end{equation}
The eigenvalues of matrix residues at singular points have the form
\begin{equation}
\mathbf{A}: \left\{0,-\frac{1}{2} (\varepsilon +1),\frac{1}{2}(\varepsilon +1)\right\},
\qquad
\mathbf{B}:\left\{0,0,-\frac{1}{2} (3+2 \varepsilon)\right\},
\qquad
\mathbf{C}:\left\{1,\frac{1}{2}(\varepsilon +1),\frac{\ep}{2}\right\}.
\end{equation}
So, according to our classification we are dealing with the case E. 
Making change of variable
\begin{equation}
z=-\frac{4 z_4^2}{(z_4^2-1)^2},
\end{equation}
and reducing the differential system to $\ep$-form with Lee algorithm we get
\begin{equation}
\frac{d}{dz_4}\Jt = \ep\left(\frac{\Mt_0}{z_4}+\frac{\Mt_1}{z_4-1}+\frac{\Mt_{-1}}{z_4+1}+\frac{z_4\Mt_w}{z_4^2+1}\right)\Jt
\label{eq:Example1PFQEpsilonForm}
\end{equation}
where
\begin{equation}
\Mt_0=
\left(
\begin{array}{ccc}
0 & 0 & 0 \\
0 & 1  & 0 \\
0 & 0 & -1  \\
\end{array}
\right),
\qquad
\Mt_1=
\left(
\begin{array}{ccc}
0 & 0 & 0 \\
-i  & 1  & 0 \\
i   & 0 & 1  \\
\end{array}
\right),
\qquad
\Mt_{-1}=
\left(
\begin{array}{ccc}
0 & 0 & 0 \\
i  & 1  & 0 \\
-i  & 0 & 1  \\
\end{array}
\right)
\end{equation}
and
\begin{equation}
\Mt_w = \left(
\begin{array}{ccc}
0 & 0 & 0 \\
0 & -4  & 0 \\
0 & 4  & 0 \\
\end{array}
\right).
\end{equation}
The boundary conditions in the transformed basis take the form
\begin{equation}
\Jt\big|_{z_4=0} = \left\{\frac{1}{8} (\varepsilon -1) (\varepsilon +1)^2,0,0\right\}.
\end{equation}
and the corresponding transformation matrix is given by
\begin{equation}
T=\left(
\begin{array}{ccc}
\frac{8}{\varepsilon ^2-1} & \frac{8 i z_4}{\varepsilon ^2-1} &
\frac{4 i \left(z_4^2-1\right)}{\left(\varepsilon ^2-1\right) z_4}
\\
-\frac{4 \varepsilon }{\varepsilon ^2-1} & -\frac{4 i z_4
	\left(\varepsilon +(\varepsilon +1) z_4^2-1\right)}{(\varepsilon
	-1) (\varepsilon +1) \left(z_4^2+1\right)} & -\frac{2 i
	\left(z_4^2-1\right)}{(\varepsilon -1) z_4} \\
\frac{2 \varepsilon  \left(\varepsilon +(\varepsilon +3) z_4^4+2
	(\varepsilon +1) z_4^2+3\right)}{\left(\varepsilon ^2-1\right)
	\left(z_4^2+1\right){}^2} & \frac{2 i z_4 P(z_4)}{\left(\varepsilon ^2-1\right)
	\left(z_4^2+1\right){}^3} & \frac{i (\varepsilon +3)
	\left(z_4^2-1\right)}{(\varepsilon -1) z_4} \\
\end{array}
\right)
\end{equation}
where
\begin{equation}
P(z_4)=\varepsilon ^2+4
\varepsilon +\left(\varepsilon ^2+4 \varepsilon +3\right)
z_4^6+\left(3 \varepsilon ^2+12 \varepsilon +5\right)
z_4^4+\left(3 \varepsilon ^2-4 \varepsilon -7\right)
z_4^2-1.
\end{equation}
The differential system in $\ep$-form \eqref{eq:Example1PFQEpsilonForm} can then be easily integrated in terms of MPLs and we get
\begin{multline}
~_3 F_2 \left(\begin{array}{c} 1,\frac{\varepsilon +1}{2},\frac{\varepsilon}{2}  \\ \frac{1-\varepsilon }{2},\frac{\varepsilon +3}{2}  \end{array} \Big|z\right)=1+\varepsilon 
\left(-\frac{\left(z_4^2+1\right) G\left(-1,z_4\right)}{2
	z_4}+\frac{\left(z_4^2+1\right) G\left(1,z_4\right)}{2
	z_4}+1\right)
\\
+\varepsilon ^2 \Bigg(\left(2 z_4+\frac{2}{z_4}\right)
G\left(f_4^1,-1,z_4\right)-\frac{2 \left(z_4^2+1\right)
	G\left(f_4^1,1,z_4\right)}{z_4}+\left(\frac{1}{2 z_4}-\frac{3
	z_4}{2}\right) G\left(0,-1,z_4\right)
\\
+\left(\frac{3z_4}{2}-\frac{1}{2 z_4}\right)
G\left(0,1,z_4\right)-\frac{\left(z_4^2+1\right)
	G\left(-1,z_4\right)}{2 z_4}+\frac{\left(z_4^2+1\right)
	G\left(1,z_4\right)}{2 z_4}-\frac{\left(z_4^2+1\right)
	G\left(-1,-1,z_4\right)}{2 z_4}
\\
+\frac{\left(z_4^2+1\right)
	G\left(-1,1,z_4\right)}{2 z_4}-\frac{\left(z_4^2+1\right)
	G\left(1,-1,z_4\right)}{2 z_4}+\frac{\left(z_4^2+1\right)
	G\left(1,1,z_4\right)}{2 z_4}\Bigg)+\OO(\ep^3).
\end{multline}
From this solution using relations \eqref{eq:DifferentialRelationsA} and \eqref{eq:DifferentialRelationsB} we can get expansions for other $~_3 F_2$ functions. For example, we have
\begin{multline}
~_3 F_2 \left(\begin{array}{c} 1,\frac{\varepsilon +1}{2},\frac{\varepsilon +2}{2}  \\ \frac{1-\varepsilon }{2},\frac{\varepsilon +3}{2}  \end{array} \Big|z\right) = \mathcal{DF}_{\frac{\varepsilon}{2}}^z  ~_3 F_2 \left(\begin{array}{c} 1,\frac{\varepsilon +1}{2},\frac{\varepsilon}{2}  \\ \frac{1-\varepsilon }{2},\frac{\varepsilon +3}{2}  \end{array} \Big|z\right) =
\frac{\left(z_4^2-1\right){}^2
	\left(G\left(-1,z_4\right)-G\left(1,z_4\right)\right)}{2
	\left(z_4^3+z_4\right)}
\\
+\frac{\varepsilon  \left(z_4^2-1\right)}{2
	\left(z_4^3+z_4\right)} \Big(-4 G\left(f_4^1,-1,z_4\right)+4
G\left(f_4^1,1,z_4\right)+G\left(-1,z_4\right)-G\left(1,z_4\right)
\\
+G\left(-1,-1,z_4\right)-G\left(-1,1,z_4\right)+G\left(1,-1,z_4\right)-G\left(1,1,z_4\right)
\\
+\frac{\left(3 z_4^2+1\right)
	G\left(0,1,z_4\right)}{1-z_4^2}+\frac{\left(3 z_4^2+1\right)
	G\left(0,-1,z_4\right)}{z_4^2-1}\Big)+\OO(\ep^2).
\end{multline}

\subsection{Example for case F}
Finally, let us consider $\ep$-expansion of
\begin{equation}
~_2 F_1 \left(\begin{array}{c} \frac{1-3\ep}{3},\frac{1+\ep}{3} \\ \frac{1}{3}  \end{array} \Big|z\right)
\end{equation}
The $\M$-matrix for differential system in this case is given by
\begin{equation}
\M = \left(
\begin{array}{cc}
0 & \frac{1}{z} \\
\frac{(\varepsilon +1) (3 \varepsilon -1)}{9 (z-1)} & \frac{2
	(\varepsilon  z-z-1)}{3 (z-1) z} \\
\end{array}
\right)
\end{equation}
The eigenvalues of matrix residues at singular points are found to be
\begin{equation}
\mathbf{A}: \left\{\frac{2}{3},0\right\},
\qquad
\mathbf{B}: \left\{0,\frac{2 (\varepsilon -2)}{3}\right\},
\qquad
\mathbf{C}: \left\{\frac{\varepsilon +1}{3},\frac{1}{3}-\varepsilon \right\}.
\end{equation}
So, now according to our classification we are dealing with case F, which satisfies additional restrictions and thus the $\ep$-expansion can be calculated
in terms of multiple polylogarithms. The general procedure in this case is to make first eigenvalues at infinity integer. To do this, we employ transformation matrix
\begin{equation}
\T = z^{-1/3}\left(
\begin{array}{cc}
1 & 0 \\
0 & 1 \\
\end{array}
\right)\, ,
\end{equation}
after which the matrix of differential system  takes the form
\begin{equation}
\M' = \left(
\begin{array}{cc}
\frac{1}{3 z} & \frac{1}{z} \\
\frac{(\varepsilon +1) (3 \varepsilon -1)}{9 (z-1)} & \frac{2
	\varepsilon  z-z-3}{3 (z-1) z} \\
\end{array}
\right), ,
\end{equation}
while the eigenvalues of matrix residues at singular points become
\begin{equation}
\mathbf{A}': \left\{1,\frac{1}{3}\right\},
\qquad
\mathbf{B}': \left\{0,\frac{2 (\varepsilon -2)}{3}\right\},
\qquad
\mathbf{C}': \left\{-\varepsilon ,\frac{\varepsilon }{3}\right\}.
\end{equation}
Now we can use the variable change 
$z \rightarrow \frac{z_1^3}{1+z_1^3}$
to make all eigenvalues integers and further apply transformation matrix
\begin{equation}
\T' = \left(
\begin{array}{cc}
1 & -\frac{3}{4} \\
-\frac{1}{3} z_1^2 \left((2 \varepsilon -1) z_1-\varepsilon \right)
& \frac{1}{4} z_1^2 \left(3 \varepsilon +(2 \varepsilon -1)
z_1\right) \\
\end{array}
\right).
\end{equation}
to finally reduce the differential system to $\ep$-form with
\begin{equation}
\Mt = 
\left(
\begin{array}{cc}
-\frac{3   \left(2 z_1^2-z_1+1\right)}{4
	\left(z_1^3+1\right)} & \frac{9   \left(2
	z_1+1\right)}{16 \left(z_1^2-z_1+1\right)} \\
\frac{  \left(2 z_1-3\right)}{3 \left(z_1^2-z_1+1\right)}
& -\frac{  \left(2 z_1^2+3 z_1-3\right)}{4
	\left(z_1^3+1\right)} \\
\end{array}
\right)
\end{equation}
The reduced differential system is easily integrated and we get
\begin{equation} 
~_2 F_1 \left(\begin{array}{c} \frac{1-3\ep}{3},\frac{1+\ep}{3} \\ \frac{1}{3}  \end{array} \Big|z\right)
=\sqrt[3]{z_1^3+1} + 
\varepsilon \sqrt[3]{z_1^3+1}  \left(\frac{2}{3} 
G\left(f_6^0,z_1\right)-\frac{4}{3} 
G\left(f_6^1,z_1\right)-\frac{2}{3} 
G\left(-1,z_1\right)\right) +\OO(\ep^2).
\end{equation}

\subsection{Overview of the algorithm}
\label{section::Algorithm}
Putting it all together let us give a summary of the proposed algorithm. As input we have a hypergeometric function $_pF_q$ with indices $a_i$ and $b_j$  linear in $\ep$ together with the desired $\ep$-expansion order $o$. And as output we want to have the $\ep$-expansion in terms of multiple polylogarithms up to specified order $o$. Then, to achieve this goal we should perform the following steps:
\begin{enumerate}
	\item First of all, optionally, we may define a new set of indices $\tilde{a}_i$ and $\tilde{b}_j$ such that $\tilde{a}_i(\ep = 0) < 1$, $\tilde{b}_i(\ep = 0) > 0$ and there exists a sequence ${s_1,\dots,s_l}$ such that
	\begin{equation}
	~_p F_q \left(
	\begin{array}{c}
	a_1,\dots,a_p  \\ b_1, \dots, b_q
	\end{array} \Big|z\right)
	=\mathcal{DF}_{s_1}^z\dots \mathcal{DF}_{s_l}^z
	~_p F_q \left(
	\begin{array}{c}
	\tilde{a}_1,\dots,\tilde{a}_p  \\ \tilde{b}_1, \dots, \tilde{b}_q
	\end{array} \Big|z\right).
	\end{equation}
	From \eqref{eq:DifferentialRelationsA} and \eqref{eq:DifferentialRelationsB} it is obvious that such sequence can always be found.
	\item For the new function $~_p F_q \left(
	\begin{array}{c}
	\tilde{a}_1,\dots,\tilde{a}_p  \\ \tilde{b}_1, \dots, \tilde{b}_q
	\end{array} \Big|z\right)$ using \eqref{eq:DifferentialEquation} and \eqref{eq:DiffEqForBasisElementsFn} we compose a system of differential equations of the form \eqref{eq:DiffEqFuchsian}. As a result at this step we get the matrix $\mathbf{M} = \frac{\mathbf{A}}{z}+\frac{\mathbf{B}}{z-1}$.
	\item Next, we calculate the eigenvalues of the obtained matrices $\mathbf{A}$, $\mathbf{B}$ and $\mathbf{C}=-\mathbf{A}-\mathbf{A}$ and use classification presented in Table \ref{tab:polylog-cases} to determine a required variable change. If the set of eigenvalues does not match any of the cases presented in Table \ref{tab:polylog-cases}, then the algorithm stops.  
	\item Now we make necessary change of variables and use Lee algorithm to bring the differential system to $\ep$-form. At the end of this step we get differential system in $\ep$-form and found expressions for $\Mt$ and $\T$ matrices.
	\item The reduced to $\ep$-form differential system is integrated in terms of multiple polylogarithms and integration constants are determined from boundary conditions. As a result we get $\ep$-expansion for the function $~_p F_q \left(
	\begin{array}{c}
	\tilde{a}_1,\dots,\tilde{a}_p  \\ \tilde{b}_1, \dots, \tilde{b}_q
	\end{array} \Big|z\right)$. 
	\item Finally, we use a sequence of differential shift operators $\mathcal{DF}_{s_1}^z\dots \mathcal{DF}_{s_l}^z$ to obtain $\ep$-expansion for the original function $~_p F_q \left(
	\begin{array}{c}
	a_1,\dots,a_p  \\ b_1, \dots, b_q
	\end{array} \Big|z\right)$.
\end{enumerate}

\section{Expansion of Appell functions}
\label{sec:AppellInG}
The approach used in previous section to obtain $\ep$-expansion of generalized hypergeometric functions of one variable can be naturally generalized to hypergeometric functions of several variables. As a first step let us consider Appell functions of two variables. The four Appell functions can be defined in terms of their series expansions \cite{bateman1953higher,Schlosser:2013hbz}
\begin{align}
F_1(\alpha,\beta_1,\beta_2,\gamma; x, y) &= \sum\limits_{m,n=0}^{\infty}\frac{(\alpha)_{m+n}(\beta_1)_m(\beta_2)_n}{(\gamma)_{m+n}m!n!}x^my^n, \qquad &|x|<1,~|y|<1,
\\
F_2(\alpha,\beta_1,\beta_2,\gamma_1, \gamma_2; x, y) &= \sum\limits_{m,n=0}^{\infty}\frac{(\alpha)_{m+n}(\beta_1)_m(\beta_2)_n}{(\gamma_1)_{m}(\gamma_2)_{n}m!n!}x^my^n, \qquad &|x|+|y|<1,
\\
F_3(\alpha_1,\alpha_2,\beta_1,\beta_2,\gamma; x, y) &= \sum\limits_{m,n=0}^{\infty}\frac{(\alpha_1)_{m}(\alpha_2)_{n}(\beta_1)_m(\beta_2)_n}{(\gamma)_{m+n}m!n!}x^my^n, \qquad &|x|<1,~|y|<1, 
\\
F_4(\alpha,\beta,\gamma_1, \gamma_2; x, y) &= \sum\limits_{m,n=0}^{\infty}\frac{(\alpha)_{m+n}(\beta)_{m+n}}{(\gamma_1)_{m}(\gamma_2)_{n}m!n!}x^my^n, \qquad &\sqrt{|x|}+\sqrt{|y|}<1
\end{align}
Similar to the case of generalized hypergeometric functions of one variable one can introduce the differential shift operators also for Appell functions. Their action is similar to \eqref{eq:DifferentialRelationsA} and \eqref{eq:DifferentialRelationsB}:
\begin{align}
\left(\frac{a}{\gamma-1}\frac{\partial}{\partial x}+\frac{b}{\gamma-1}\frac{\partial}{\partial y}+1\right)\sum\limits_{m,n=0}^{\infty}\frac{f(m,n)x^my^n}{(\gamma)_{am+bn}} & = \sum\limits_{m,n=0}^{\infty}\frac{f(m,n)x^my^n}{(\gamma-1)_{am+bn}}
\label{eq:DFForAppel1}
\\
\left(\frac{a}{\alpha}\frac{\partial}{\partial x}+\frac{b}{\alpha}\frac{\partial}{\partial y}+1\right)\sum\limits_{m,n=0}^{\infty}(\alpha)_{am+bn}f(m,n)x^my^n & = \sum\limits_{m,n=0}^{\infty}(\alpha+1)_{am+bn}f(m,n)x^my^n
\label{eq:DFForAppel2}
\end{align}
and they also can be used to reduce the eigenvalues of matrix residues in the corresponding differential systems prior to using balance transformations. To write down the required matrix differential systems we need to decide which basis functions to use. The natural choice is the function itself together with its partial derivatives. Also, it is enough to have a differential system only in one variable, for example $x$. This is due to the fact, that Appell functions degenerate into ordinary hypergeometric functions when one of their arguments is zero. Thus, we do not need a complete Pfaffian system here, that is the full differential of basis functions, and may proceed with only one of its components. Boundary conditions for differential system in one variable can be obtained using the method of previous section. This is especially convenient considering the fact that by solving the differential system we immediately obtain solutions for hypergeometric function together with its derivatives from which we can easily assemble the boundary conditions for the Appell differential system.

As a simple example, consider the Appell function $F_1$. The latter satisfies a system composed of two second-order partial differential equations
\begin{align}
\left[x(1-x)\frac{\partial^2}{\partial x^2} + y(1-x)\frac{\partial^2}{\partial x \partial y} + \left[\gamma - (\alpha +\beta_1+1)x\right]\frac{\partial}{\partial x}-\beta_1 y \frac{\partial}{\partial y} - \alpha \beta_1\right]F_1 & = 0,
\label{eq:F1Partial1}
\\
\left[y(1-y)\frac{\partial^2}{\partial y^2} + x(1-y)\frac{\partial^2}{\partial x \partial y} + \left[\gamma - (\alpha +\beta_2+1)y\right]\frac{\partial}{\partial y}-\beta_2 x \frac{\partial}{\partial x} - \alpha \beta_2\right]F_1 & = 0.
\label{eq:F1Partial2}
\end{align}
This system of equations has three linearly independent solutions. Accordingly, it can be reduced to a system of three linear differential equations in one of the variables with second variable considered as a parameter. Choosing the function basis as
\begin{equation}
\J_1  = \left\{F_1, x\frac{\partial}{\partial x}F_1, y\frac{\partial}{\partial y}F_1 \right\},
\label{eq:F1Basis}
\end{equation}
it is logical to assume that these three components will be linearly independent.
Then from the partial differential equations \eqref{eq:F1Partial1}-\eqref{eq:F1Partial2} it follows that chosen basis functions satisfy the following matrix differential system
\begin{equation}
\label{eq:DiffEqFuchsianF1}
\frac{\partial}{\partial x}\J_1 = \left(\frac{\mathbf{A}_0}{x}+\frac{\mathbf{A}_1}{x-1}+\frac{\mathbf{A}_y}{x-y}\right)\J_1\, ,
\end{equation}
where
\begin{equation}
\mathbf{A}_0 = \left(
\begin{array}{ccc}
0 & 1 & 0 \\
0 & \beta_2-\gamma +1 & 0 \\
0 & -\beta_2 & 0 \\
\end{array}
\right),
\mathbf{A}_1 =\left(
\begin{array}{ccc}
0 & 0 & 0 \\
-\alpha\beta_1 & -\alpha -\beta_1+\gamma 
-1 & -\beta_1 \\
0 & 0 & 0 \\
\end{array}
\right),
\mathbf{A}_y = \left(
\begin{array}{ccc}
0 & 0 & 0 \\
0 & -\beta_2 & \beta_1 \\
0 & \beta_2 & -\beta_1 \\
\end{array}
\right).
\end{equation}
and matrix residues eigenvalues at singular points have the form
\begin{align}
\mathbf{A}_0: & \{0, 0,   \beta _2-\gamma +1 \},
\\
\mathbf{A}_1:& \{0,   0,   -\alpha -\beta _1+\gamma -1 \},
\\
\mathbf{A}_y:& \{0,   0,   -\beta _1-\beta _2 \}.
\\
\mathbf{A}_{\infty}: & \{\alpha ,  \beta_1,   \beta_1 \}.
\end{align}
The boundary conditions are given by the usual hypergeometric function $$F_1(\alpha,\beta_1,\beta_2,\gamma; 0, y) = ~_2 F_1 \left(\begin{array}{c} \alpha,\beta_2 \\ \gamma \end{array} \Big|y\right)$$ together with its first derivative over $y$. The $\ep$-expansions for the latter can then be found by the application of the algorithm from the previous section. As already noted, solving the corresponding differential system gives us solutions for both $_2 F_1$ and $y\frac{d}{dy}~_2 F_1$ functions.

Let us now see how the differential system \eqref{eq:DiffEqFuchsianF1} is obtained in more detail. Here, we will restrict ourselves to the use of ordinary linear algebra and for more general procedure for reducing a system of equations to Pfaffian form with the use of the Gröbner basis techniques we refer the interested reader to \cite{Koutschan2010}. Denoting basis functions as $\J_1 = \left\{J_a, J_b, J_c \right\}$ the equations \eqref{eq:F1Partial1} and \eqref{eq:F1Partial2} take the form
\begin{equation}
x \left((x-1)
\left(J'_b+J'_c\right)+\beta_1
J_c\right)+J_b (-\gamma +x (\alpha
+\beta_1)+1)+\alpha  \beta_1 x J_a = 0,
\label{eq:F1Partial1Not}
\end{equation}
\begin{equation}
(y-1) y^2 \frac{\partial^2}{\partial y^2}F_1+x (y-1) J'_c+J_c (y
(\alpha +\beta_2+1)-\gamma )+\alpha
\beta_2 J_a y+\beta_2 y J_b
=0,
\label{eq:F1Partial2Not}
\end{equation}
where the prime stands for the derivative with respect to $x$. In addition we have one extra equation that follows from the definition of basis functions
\begin{equation}
x J'_a-J_b = 0.
\label{eq:F1Partial0Not}
\end{equation}
In this notation it becomes immediately clear that the original equations have an "extra" term $\frac{\partial^2}{\partial y^2}F_1$ that needs to be expressed through our basis functions  $\left\{J_a, J_b, J_c \right\}$. To achieve this we differentiate equation \eqref{eq:F1Partial1} with respect to $y$ and equation  \eqref{eq:F1Partial2} with respect to $x$ to get
\begin{equation}
-\beta_1 y^2 \frac{\partial^2}{\partial y^2}F_1-(x-1) y
\left(y \frac{\partial^3}{\partial x \partial y^2}F_1+x
\frac{\partial^3}{\partial x^2 \partial y}F_1\right)+J'_c (\gamma -x
(\alpha +\beta_1+2)+1)-(\alpha +1)
\beta_1 J_c = 0
\end{equation}
\begin{multline}
x (y-1) y \left(y \frac{\partial^3}{\partial x \partial y^2}F_1+x
\frac{\partial^3}{\partial x^2 \partial y}F_1\right)
+x J'_c (-\gamma +y
(\alpha +\beta_2+2)-1)+\beta_2 x y J'_b+\alpha  \beta_2 y J_b=0
\end{multline}
Next, we note, that in addition to the basis functions $\left\{J_a, J_b, J_c \right\}$ and their first derivatives with respect to $x$, these equations depend on two combination $\left(y \frac{\partial^3}{\partial x \partial y^2}F_1+x
\frac{\partial^3}{\partial x^2 \partial y}F_1\right)$ and $\frac{\partial^2}{\partial y^2}F_1$. Therefore, these equations allow us to expresses mentioned combinations solely in terms of the basis functions. We are interested only in one combination, for which we get
\begin{multline}
(y-1) y^2 \frac{\partial^2}{\partial y^2}F_1 = \frac{1}{\beta_1 x}  \Big[x J'_c (x (\alpha +\beta_1-\gamma
-\beta_1 y+\beta_2 y+1)-y
(\alpha +\beta_2-\gamma
+1))
\\
+\beta_2 (x-1) x y J'_b
-(\alpha
+1) \beta_1 J_c x (y-1)+\alpha 
\beta_2 (x-1) y J_b \Big].
\end{multline}
Finally, substituting this expression back into equation \eqref{eq:F1Partial2Not} and combining together equations  \eqref{eq:F1Partial1Not},  \eqref{eq:F1Partial2Not} and  \eqref{eq:F1Partial0Not} we get the desired differential system \eqref{eq:DiffEqFuchsianF1}.

With other Appell functions the situation is similar, the only difference is that now their systems of equations have four linearly independent solutions. Accordingly, we will choose the bases for their matrix differential systems as\footnote{We have also experimented with other functions bases $\left\{F_i,~ x\frac{\partial}{\partial x}F_i,~ y\frac{\partial}{\partial y}F_i, ~y^2\frac{\partial^2}{\partial y }F_i \right\},~ i=2,3$ and $\left\{F_4,~ x\frac{\partial}{\partial x}F_4,~ y\frac{\partial}{\partial y}F_4, ~yx\frac{\partial^2}{\partial y\partial x }F_4 \right\}$, but they turned out to be less effective, especially for complex problems.}
\begin{equation}
\J_i  = \left\{F_i,~ x\frac{\partial}{\partial x}F_i,~ y\frac{\partial}{\partial y}F_i, ~yx\frac{\partial^2}{\partial y\partial x }F_i \right\},\qquad i=2,3
\end{equation}
and
\begin{equation}
\J_4  = \left\{F_4,~ x\frac{\partial}{\partial x}F_4, ~y\frac{\partial}{\partial y}F_4,~ y^2\left(x-(1-\sqrt{y})^2\right)\frac{\partial^2}{\partial y^2}F_4 +y\left(x-(1-\sqrt{y})^2-1\right)\frac{\partial}{\partial y}F_4 \right\}.
\label{eq:F4Basis}
\end{equation}
In the case of $F_4$ function such a non-trivial basis was chosen to simplify the eigenvalues of the matrix residues in advance. This is an optional choice, but it allows to significantly speed up calculations. In fact, this is the function basis obtained from  $\left\{F_4,~ x\frac{\partial}{\partial x}F_4,~ y\frac{\partial}{\partial y}F_4, ~y^2\frac{\partial^2}{\partial y }F_4 \right\}$ with the balance transformation. The matrix differential systems in these bases are given by
\begin{equation}
\label{eq:DiffEqFuchsianF2}
\frac{d}{dx}\J_2 = \left(\frac{\mathbf{B}_0}{x}+\frac{\mathbf{B}_1}{x-1}+\frac{\mathbf{B}_{y}}{x-1+y}\right)\J_2
\end{equation}
\begin{equation}
\label{eq:DiffEqFuchsianF3}
\frac{d}{dx}\J_3 = \left(\frac{\mathbf{C}_0}{x}+\frac{\mathbf{C}_1}{x-1}+\frac{\mathbf{C}_{y}}{x-\frac{y}{y-1}}\right)\J_3
\end{equation}
\begin{equation}
\label{eq:DiffEqFuchsianF4}
\frac{d}{dx}\J_4 = \left(\frac{\mathbf{D}_0}{x}+\frac{\mathbf{D}_{y}}{x-(\sqrt{y}+1)^2}+\frac{\mathbf{D}_{-y}}{x-(\sqrt{y}-1)^2}\right)\J_4
\end{equation}
To save space we will not write down here explicit expressions for matrices $\mathbf{B}$,  $\mathbf{C}$ and $\mathbf{D}$, only note that they are in general $y$-dependent. However, it is useful to know explicit expressions for matrix residues eigenvalues at singular points. For $F_2$ function they are given by
\begin{align}
\mathbf{B}_0: & \left\{0,0,1-\gamma _1,1-\gamma _1\right\},
\\
\mathbf{B}_1:& \left\{0,0,0,-\alpha -\beta _1+\beta _2+\gamma _1-1\right\},
\\
\mathbf{B}_y:& \left\{0,0,0,-\alpha -\beta _1-\beta _2+\gamma _1+\gamma _2-2\right\},
\\
\mathbf{B}_{\infty}: & \left\{\alpha ,\beta _1,\beta _1,\alpha -\gamma _2+1\right\}\, ,
\end{align}
for $F_3$ function by
\begin{align}
\mathbf{C}_0: & \left\{0,0,\alpha _2-\gamma +1,\beta _2-\gamma +1\right\},
\\
\mathbf{C}_1:& \left\{0,0,0,-\alpha _1-\beta _1+\gamma -1\right\},
\\
\mathbf{C}_y:& \left\{0,0,0,-\alpha _1-\alpha _2-\beta _1-\beta _2+\gamma -1\right\},
\\
\mathbf{C}_{\infty}: & \left\{\alpha _1,\alpha _1,\beta _1,\beta _1\right\}\, 
\end{align}
and for $F_4$ function we have
\begin{align}
\mathbf{D}_0: & \left\{0,0,1-\gamma _1,1-\gamma _1\right\},
\\
\mathbf{D}_{y}:& \left\{0,0,0,-\alpha -\beta +\gamma _1+\gamma _2-\frac{5}{2}\right\},
\\
\mathbf{D}_{-y}:& \left\{0,0,0,-\alpha -\beta +\gamma _1+\gamma _2-\frac{3}{2}\right\},
\\
\mathbf{D}_{\infty}: & \left\{\alpha ,\beta ,\alpha -\gamma _2+1,\beta -\gamma _2+1\right\}.
\end{align}
In general, the algorithm for the $\ep$-expansion of Appell functions works in the same way as in the case of $_pF_q$ functions. However, the presence of an additional parameter complicates the calculations and the search for a suitable variable change becomes especially difficult in the general. Thus, for the moment, we will restrict ourselves only to two cases: the one when all eigenvalues of matrix residues are integers (Case A) and another one when two matrix residues have half-integer\footnote{In general, similar to the case of $\pFq$ function, one can consider eigenvalues with common denominator greater than two and this is implemented in the program. However, this option does not always work stable enough and should be considered as experimental.} eigenvalues (Case B). In the latter case the change of variable has the following form
\begin{align}
x_{\rm new} & =\sqrt{\frac{x-x_1}{x-x_2}},\qquad x_{1,2} \ne \infty
\\
x_{\rm new} & =\sqrt{x-x_1},\qquad x_{2} = \infty
\end{align}
where $x_1$ and $x_2$ are two singular points at which the eigenvalues of the matrix residues have half-integer values. With more complex cases, such as half-integer eigenvalues at three singular points, there are computational problems which require different approaches both for balancing eigenvalues and search for a suitable variable change. We consider this as a topic for future research. It is interesting to note that for functions $F_1$, $F_2$ and $F_3$ both A and B cases are possible, while for the $F_4$ function only case B is possible even if all indices are integers.

Here are examples of  $\ep$-expansion of some Appell functions which can be expanded in terms of multiple polylogarithms and require variable change\footnote{These are the functions which appear in the calculation of some one-loop Feynman diagrams \cite{Tarasov:2022clb}.}: 
\begin{multline}
F_1\left(\frac{1}{2};1,2-\frac{d}{2};\frac{3}{2};x,y\right) =\frac{G\left(-1,\sqrt{x}\right)-G\left(1,\sqrt{x}\right)}{2 \sqrt{x}}+ \frac{\varepsilon}{2\sqrt{x}}\Big(-G\left(-1,\sqrt{x}\right)
\left(G\left(-1,\sqrt{y}\right)+G\left(1,\sqrt{y}\right)\right)
\\
+G\left(1,\sqrt{y}\right)
\left(G\left(-\sqrt{y},\sqrt{x}\right)-G\left(\sqrt{y},\sqrt{x}\right)+G\left(1,\sqrt{x}\right)\right)
\\
+G\left(-1,\sqrt{y}\right)
\left(-G\left(-\sqrt{y},\sqrt{x}\right)+G\left(\sqrt{y},\sqrt{x}\right)+G\left(1,\sqrt{x}\right)\right)
\\
-G\left(-\sqrt{y},-1,\sqrt{x}
\right)+G\left(-\sqrt{y},1,\sqrt{x}\right)-G\left(\sqrt{y},-1,\sqrt{x}\right)
\\
+G\left(\sqrt{y},1,\sqrt{x}\right)+2
G\left(0,-1,\sqrt{x}\right)-2 G\left(0,1,\sqrt{x}\right)\Big)+\OO(\ep^2),
\end{multline}
\begin{multline}
F_1\left(\frac{d-2}{2};1,\frac{1}{2};\frac{d}{2};x,y\right) = 
\frac{1}{x}\sqrt{\frac{x}{x-y}}
\left(G\left(1,\sqrt{\frac{x}{x-y}}\right)-G\left(-1,\sqrt{\frac{x}{x-y}}\right)\right.
\\
\left.+G\left(-\frac{1}{\sqrt{1-y}},\sqrt{\frac{x}{x-y}}\right)-G\left(\frac{1}{\sqrt{1-y}},\sqrt{\frac{x}{x-y}}\right)\right)+\OO(\ep)
\end{multline}
and
\begin{multline}
F_3\left(1,1,1,\frac{d-3}{2},\frac{d}{2},x,y\right) = \frac{1}{x}\sqrt{\frac{x}{x+y-xy}}
\left[G\left(1,\sqrt{\frac{x}{x-\frac{y}{y-1}}}\right)-G\left(-1,\sqrt{\frac{x}{x-\frac{y}{y-1}}}\right)\right.
\\
\left.+G\left(-\sqrt{1-y},\sqrt{\frac{x}{x-\frac{y}{y-1}}}\right)-G\left(\sqrt{1-y},\sqrt{\frac{x}{
		x-\frac{y}{y-1}}}\right)\right]+\OO(\ep)
\end{multline}
where $d = 4-2\ep$. Higher $\ep$-correction terms can be obtained with the help  of our package and can be found in the provided Mathematica notebook with examples.

\subsection{Example}
As a particular example of the calculation, consider a function
\begin{equation}
F_1(1,\ep,\frac{1}{2},2-\frac{2\ep}{3}; x, y)\, ,
\end{equation}
which provides a fairly clear example of the main points present in general calculation. The matrix of corresponding differential system in the basis \eqref{eq:F1Basis} has the form
\begin{equation}
\M = \left(
\begin{array}{ccc}
0 & \frac{1}{x} & 0 \\
-\frac{\varepsilon }{x-1} & \frac{-\left((\varepsilon +1)
	x^2\right)+x \left(-\frac{2 \varepsilon }{3}+\left(\varepsilon
	+\frac{1}{2}\right) y+1\right)+\left(\frac{2 \varepsilon
	}{3}-\frac{1}{2}\right) y}{(x-1) x (x-y)} & \frac{\varepsilon 
	(y-1)}{(x-1) (x-y)} \\
0 & \frac{y}{2x \left(x- y\right)} & -\frac{\varepsilon }{x-y} \\
\end{array}
\right)
\end{equation}
and the boundary conditions are given by the vector
\begin{equation}
\J_b= \left\{~_2 F_1 \left(
\begin{array}{c}
1, \frac{1}{2}  \\ 2-\frac{2\ep}{3}
\end{array} \Big|y\right),
0,
y\frac{d}{dy}~_2 F_1 \left(
\begin{array}{c}
1, \frac{1}{2}  \\ 2-\frac{2\ep}{3}
\end{array} \Big|y\right)
\right\}.
\end{equation}
The $\ep$-expansion of the latter can be easily performed in terms of multiple polylogarithms using the algorithm from the previous section. The set of eigenvalues for $\M$-matrix residues at singular points is given by 
\begin{align}
0: & \left\{ 0 ~~ 0 ~~ -\frac{1}{2}+\frac{2\ep}{3} \right\},
\\
1:& \left\{ -\frac{5 \varepsilon }{3} ~~ 0 ~~ 0 \right\},
\\
y:& \left\{ 0 ~~ 0 ~~ -\varepsilon -\frac{1}{2} \right\}.
\\
\infty: & \left\{ 1 ~~ \varepsilon  ~~ \varepsilon \right\}.
\end{align}
We see that there are two half-integer eigenvalues at the points $x=0$ and $x = y$. Therefore, they can be made integer by performing a transformation to a new variable $z_{13} = \sqrt{x/(x-y)}$. After that, the differential system can be reduced to $\ep$-form using the usual balance transformations. This way we get
\begin{equation}
\frac{\partial \Jt}{\partial z_{13}} = \ep \Mt \Jt 
\end{equation}
where
\begin{equation}
\Mt = 
\left(
\begin{array}{ccc}
\frac{2 \left(3 z_3^2 z_{13}^4-5 z_{13}^2+2\right)}{3 z_{13} \left(z_3^2
	z_{13}^2-1\right) \left(z_{13}^2-1\right)} & \frac{22 z_3^2
	z_{13}^2}{7 \left(z_{13}^2-1\right) \left(z_3^2 z_{13}^2-1\right)}
& -\frac{11 \left(3 z_3^2 z_{13}^2-7\right)}{21
	\left(z_{13}^2-1\right) \left(z_3^2 z_{13}^2-1\right)} \\
\frac{4 (3 y+4)}{33 \left(z_3^2 z_{13}^2-1\right)} & -\frac{2 (3
	y+4) z_{13}}{7 \left(z_{13}^2-1\right) \left(z_3^2
	z_{13}^2-1\right)} & -\frac{4 (3 y+4) z_{13}}{21
	\left(z_{13}^2-1\right) \left(z_3^2 z_{13}^2-1\right)} \\
-\frac{8 z_3^2}{11( z_3^2 z_{13}^2-1)} & \frac{12 z_3^2 z_{13}}{7
	\left(z_{13}^2-1\right) \left(z_3^2 z_{13}^2-1\right)} & \frac{8
	z_3^2 z_{13}}{7 \left(z_{13}^2-1\right) \left(z_3^2
	z_{13}^2-1\right)} \\
\end{array}
\right)
\end{equation}
and $z_3 = \sqrt{1-y}$. The new function basis $\Jt$ is related to the old one through the transformation matrix $\T$ ($\J = \T\cdot\Jt$): 
\begin{equation}
\T = \left(
\begin{array}{ccc}
\frac{6}{11 z_{13}} & 0 & -1 \\
\frac{4 \varepsilon +(3-10 \varepsilon ) z_{13}^2-3}{11
	z_{13}} & 0 & \varepsilon  \\
\frac{(10 \varepsilon -3) z_{13}^2-3}{11 z_{13}} &
\frac{1}{7} (10 \varepsilon -3) & -\frac{5}{7}
(\varepsilon -1) \\
\end{array}
\right).
\end{equation}
Finally, integrating the resulting differential system, with account of  boundary conditions, we get
\begin{multline}
F_1(1,\ep,\frac{1}{2},2-\frac{2\ep}{3}; x, y)  =\frac{2}{z_3+1}+\frac{2 \varepsilon}{3 y}  \Bigg(\frac{3}{z_3}
\left(-y-\frac{z_3}{z_{13}}+1\right)
G\left(\frac{1}{z_3},z_{13}\right)-4 z_3
G\left(-1,z_3\right)
\\
+4 z_3 G\left(0,z_3\right)
+3\left(\frac{1}{z_{13}}-z_3\right) G\left(1,z_{13}\right)+3
\left(z_3+\frac{1}{z_{13}}\right)
G\left(-\frac{1}{z_3},z_{13}\right)
\\
-3\left(z_3 +\frac{1}{z_{13}}\right) G\left(-1,z_{13}\right)+4 z_3 (\log
(2)-2)+8\Bigg)+\OO(\ep^2).
\end{multline}
Also, we can use differential  shift operators to get results for more complex functions. For example, using the relation
\begin{equation}
F_1(1,1+\ep,\frac{1}{2},2-\frac{2\ep}{3}; x, y) = \left(\frac{1}{\ep}\frac{\partial}{\partial x} + 1\right) F_1(1,\ep,\frac{1}{2},2-\frac{2\ep}{3}; x, y) 
\end{equation}
and taking derivatives of $G$-functions with \eqref{eq:GtotalDifferential}
we get
\begin{multline}
F_1(1,1+\ep,\frac{1}{2},2-\frac{2\ep}{3}; x, y)=\frac{1}{x}\left(
z_{13}
G\left(1,z_{13}\right)+z_{13}
G\left(-\frac{1}{z_3},z_{13}\right)-z_{13}
G\left(\frac{1}{z_3},z_{13}\right)\right)+\OO(\ep).
\end{multline}

\subsection{Reduction formulas}
\label{section::ReductionAppell}
Sometimes, for certain values of the parameters, the Appell functions can be reduced to simpler hypergeometric functions. In addition to the trivial cases when one of the arguments or indices is zero, we have the following reduction relations\cite{bateman1953higher,Schlosser:2013hbz}
\begin{equation}
F_2(\alpha;\beta_1,\beta_2;\beta_1,\gamma_2;x,y) = (1-x)^{-\alpha}~_2F_1\left(
\begin{array}{c}
\alpha,\beta_2  \\ \gamma_2,
\end{array} \Big|\frac{y}{1-x}\right),
\end{equation}
\begin{equation}
F_2(\alpha;\beta_1,\beta_2;\gamma_1,\alpha ;x,y) = (1-y)^{-\beta_2}F_1(\beta_1;\alpha-\beta_2,\beta_2;\gamma_1;x, x/(1-y)),
\end{equation}
\begin{equation}
F_3(\alpha,\gamma-\alpha;\beta,\gamma-\beta;\gamma;x,y) = (1-y)^{\alpha+\beta-\gamma}~_2F_1\left(
\begin{array}{c}
\alpha,\beta  \\ \gamma,
\end{array} \Big|x+y-xy\right),
\end{equation}
\begin{equation}
F_3(\alpha,\gamma-\alpha;\beta_1,\beta_2;\gamma;x,y/(y-1)) = (1-y)^{\beta_2}F_1(\alpha;\beta_1,\beta_2;\gamma;x,y),
\end{equation}
\begin{equation}
F_4(\alpha;\beta;\gamma,1+\alpha+\beta-\gamma; x(1-y), y(1-x)) = 
~_2F_1\left(
\begin{array}{c}
\alpha,\beta  \\ \gamma,
\end{array} \Big|x\right)
~_2F_1\left(
\begin{array}{c}
\alpha,\beta  \\ 1+\alpha+\beta-\gamma,
\end{array} \Big|y\right),
\end{equation}
\begin{multline}
F_4(\alpha;\beta;\gamma,\beta; x(1-y), y(1-x)) = 
\\
=(1-x)^{-\alpha}(1-y)^{-\alpha}F_1\left(\alpha;1+\alpha-\gamma, \gamma- \beta;\gamma;\frac{xy}{(1-x)(1-y)},\frac{x}{x-1}\right)
\end{multline}
and similar ones obtained by permutation of indices. We do not include relations for the $F_1$ function here since they are already built in into the Wolfram Mathematica system. Obviously, these relations together with the described expansion procedure give us additional possibilities for polylogarithmic expansions in these particular cases.

\section{Expansion of Lauricella functions}
\label{sec:LauricellaInG}
A further generalization of the Appell functions to the case of more variables are  Lauricella functions \cite{bateman1953higher,Schlosser:2013hbz}. We will consider the following three functions\footnote{We are not considering  $F_C^{(n)}$ function here. The latter in general appear in the  calculations of multi-loop (hyper)elliptic diagrams such as sunsets with arbitrary masses of propagators \cite{Berends:1993ee,Lee:2019lsr}. The expansion for such integrals is beyond the scope of the present paper. Also, function $F_4$, which is a special case of the function  $F_C^{(n)}$ with $n = 2$, already presents significant computational complexity and requires a customized basis \eqref{eq:F4Basis} to be used. And it is still unclear how to choose a good basis in general case.}
\begin{equation}
F_A^{(n)}(\alpha; \beta_1,\dots, \beta_n;\gamma_1,\dots,\gamma_n; x_1,\dots,x_n) = \sum\limits_{m_1,\dots, m_n = 0}^{\infty}\frac{(\alpha)_{m_1+\dots+m_n}(\beta_1)_{m_1}\dots(\beta_n)_{m_n}}{(\gamma_1)_{m_1}\dots(\gamma_n)_{m_n}m_1!\dots m_n!}x_1^{m_1}\dots x_n^{m_n},
\end{equation}
\begin{equation}
F_B^{(n)}(\alpha_1,\dots,\alpha_n; \beta_1,\dots, \beta_n;\gamma; x_1,\dots,x_n) = \sum\limits_{m_1,\dots, m_n = 0}^{\infty}\frac{(\alpha_1)_{m_1}\dots (\alpha_n)_{m_n}(\beta_1)_{m_1}\dots(\beta_n)_{m_n}}{(\gamma)_{m_1+\dots+m_n}m_1!\dots m_n!}x_1^{m_1}\dots x_n^{m_n},
\end{equation}
\begin{equation}
F_D^{(n)}(\alpha; \beta_1,\dots, \beta_n;\gamma; x_1,\dots,x_n) = \sum\limits_{m_1,\dots, m_n = 0}^{\infty}\frac{(\alpha)_{m_1+\dots+m_n}(\beta_1)_{m_1}\dots(\beta_n)_{m_n}}{(\gamma)_{m_1+\dots+m_n}m_1!\dots m_n!}x_1^{m_1}\dots x_n^{m_n}.
\end{equation}
At $n=2$ Lauricella functions obviously reduce to Appell functions
\begin{equation}
F_A^{(2)} = F_2, \qquad F_B^{(2)} = F_3, \qquad F_D^{(2)} = F_1.
\end{equation}
All steps in the calculation of $\ep$-expansion of Lauricella functions are similar to what we did in the case of Appell functions. Index shift operators can be introduced in exactly the same way as in equations \eqref{eq:DFForAppel1} and \eqref{eq:DFForAppel2}. The derivation of differential systems also goes similarly. The boundary conditions for  functions with $n$ variables are given by functions with $(n-1)$ variables. The only important difference is the number of variables treated as parameters in differential systems, which in fact increases only computational complexity. For this reason we will skip a detailed description of the expansion procedure and limit ourselves to the discussion of its differences from that for Appell functions.

By analogy with  Appell functions the function basis can be chosen as
\begin{equation}
\left\{\theta_{x_{j_1}} \dots \theta_{x_{j_k}}F_i~\Big| ~ 0 \le k \le n,~ j_1 < j_2 < \dots < j_k\right\}, \qquad i = {A,B},
\end{equation}
and
\begin{equation}
\left\{F_D,~\theta_{x_{j}}F_D~\Big|~ j={1,\dots,n}\right\},
\end{equation}
where $\theta_a = \partial/\partial a$.
Thus, for $F_A^{(n)}$ and $F_B^{(n)}$ functions the function basis will consist from $2^n$ elements, while for simpler $F_D^{(n)}$ function the basis contains $n + 1$ elements. Also, the differential systems will now have more singular points. For $n=3$ the differential systems with respect to $x_1$ variable will have the following singularities:
\begin{align}
F_A^{(3)}: & \{0, 1, 1- x_2,1-x_3,1-x_2-x_3,\infty\},
\\
F_B^{(3)}:& \left\{0,1,\frac{x_2}{x_2-1},\frac{x_3}{x_3-1},\frac{x_2 x_3}{x_2 x_3-x_2-x_3},\infty
\right\},
\\
F_D^{(3)}:& \{0,1,x_2,x_3,\infty \}
\end{align}
The higher $n$, the more difficult it will be to obtain desired expansions. In practice, we can more or less stably obtain solutions for $n = 3$ and in some simple cases for $n = 4$. Of course, the calculation of functions $F_D^{(n)}$ is simpler than others due to the smaller basis, which grows only linearly with $n$.

\subsection{Example}

In the case of Lauricella functions the calculations become quite cumbersome. Therefore, we will only give a fairly simple example with a small amount of detail. Consider a function
\begin{equation}
F_D^{(3)}\left(\frac{1}{2}-\ep;1,\ep,\ep;1+2\ep;x,y,z\right)
\end{equation}
After choosing basis as indicated above, the $\M$-matrix in the corresponding differential system can be written as
\begin{equation}
\M = \left(
\begin{array}{cccc}
0 & \frac{1}{x} & 0 & 0 \\
\frac{1-2 \varepsilon }{2(1- x)} & \frac{2 \varepsilon 
	\left(x^2-2 x (y+z-1)+y (3 z-1)-z\right)-3(x-y) (x-z)}{2 (x-1)(x-y) (x-z)}
& \frac{y-1}{(x-1) (x-y)} & \frac{z-1}{(x-1) (x-z)} \\
0 & \frac{\varepsilon  y}{x(x- y)} & \frac{1}{y-x} & 0 \\
0 & \frac{\varepsilon  z}{x(x- z)} & 0 & \frac{1}{z-x} \\
\end{array}
\right)
\end{equation}
The set of eigenvalues of matrix residues at singular points is then easily found to be
\begin{align}
0: & \{0,0,0,0\},
\\
1: & \left\{0,0,0,3 \varepsilon -\frac{3}{2}\right\},
\\
y: & \{0,0,0,-\varepsilon -1\},
\\
z: & \{0,0,0,-\varepsilon -1\},
\\
\infty: & \left\{1,1,1,\frac{1}{2}-\varepsilon \right\}.
\end{align}
These eigenvalues can be converted to integers by a simple variable change $z_x = \sqrt{1-x}$. After which the differential system can be reduced to $\ep$-form using the usual Lee algorithm. The reduced differential system is solved straightforwardly and the desired $\ep$-expansion takes the form 
\begin{multline}
F_D^{(3)}\left(\frac{1}{2}-\ep;1,\ep,\ep;1+2\ep;x,y,z\right)=
\frac{1}{z_x}- \frac{2\ep}{z_x} \Big[
G\left(-z_y,z_x\right)
+G\left(-z_z,z_x\right)+
G\left(-1,z_y\right)
\\
-G\left(-z_y,1\right)
+ G\left(-1,z_z\right)-
G\left(-z_z,1\right)-3 \log z_x-\log
(4)\Big]+\OO(\ep^2)
\end{multline}
where $z_a = \sqrt{1-a}$
 
\section{Conclusion}
\label{section::Сonclusions}

In the present work we have studied the $\ep$-expansion of different hypergeometric functions both of one and many variables with indices linear dependent on $\ep$. In particular we were interested in cases when such expansion is expressible in terms of multiple polylogarithms. The proposed expansion procedure is based on the reduction of corresponding differential systems to $\ep$-form. In the case of generalized hypergeometric functions of one variable we have found and classified quite a lot of cases when it is possible. Still, there  may be extra exotic cases which we missed. We have reasons to believe that in all cases when the expansion in terms of the polylogarithms is possible one can devise a systematic procedure for finding a required variable change. Also, in cases when it is not possible one can still systematically perform $\ep$-expansion of generalized hypergeometric functions in terms of iterated integrals with algebraic kernels. As for the hypergeometric functions of many variables we have only touched the subject by considering $\ep$-expansion of Appell and Lauricella functions in several simple cases. The systematic study of these and other hypergeometric functions with many variables is certainly required. The current problems with the performance issues of the \texttt{Diogenes} package in applications to hypergeometric functions with many variables can also be solved. There are many directions to improve, like optimization of the reduction algorithm, parallelization and so on.  All these problems will be the subject of our future research.

\section*{Acknowledgments}
We would like to thank V.V.Bytev, R.N.Lee, A.V.Kotikov and O.L.Veretin for interesting and stimulating discussions. The work was supported by the Russian Science Foundation, grant 20-12-00205.

\appendix

\section{Multiple polylogarithms}
\label{appendix::MPLs}

The expansion coefficients of hypergeometric functions studied in the present paper were expressed in terms of so called Goncharov multiple polylogarithms\footnote{See also \cite{MPLsHopf1} for general introduction.} (MPLs) \cite{goncharov2,goncharov3}. The later are defined recursively as: 
\begin{equation}
\label{MPL_Def}
G(a_1,...,a_n;x)=\int\limits_0^x \frac{G(a_2,...,a_n;x')}{x'-a_1}dx', \qquad n>0,
\end{equation}
where $a_i,x \in \mathbb{C}$ and  $n \in \mathbb{N}$ is referred to as 
the polylogarithm weight. The recursion starts with $G(;x)=1$ and for zero indexes of polylogarithm one employs the following regularization rule
\begin{equation}
G(\underbrace{0,\dots,0}_n;x)=\frac{\log^n x}{n!}.
\end{equation}
This definition is most convenient for practical application and is the most common in particle physics. 

MPLs are well studied class of functions and the detail discussion of their properties goes far beyond the scope of present paper. More detailed overview of the MPLs properties including their Hopf algebra structure can be found in \cite{MPLsHopf1, Duhr1, Duhr2}. Here we will only mention that MPLs form a closed space under taking primitives (integration) and derivatives. If polylogarithm indexes $a_i$ are $x$ independent and $R(x)$ is some rational function of variable $x$ then the primitive of the product $R(x)\cdot G(\vec{a};x)$ can be expressed as a linear combination of some other MPLs with rational coefficients. Similarly, the derivative of $G(\vec{a}(x);f(x))$ can also be expressed as a linear combination of MPLs. For example, the total differential is written as

\begin{equation}
d G(a_1,...,a_n;a_0)=\sum\limits_{i=1}^nG(a_1,...,a_{i-1},a_{i+1},...,a_n;a_0)d \log \left(\frac{a_{i-1}-a_{i}}{a_{i+1}-a_i}\right)
\label{eq:GtotalDifferential}
\end{equation} 
where $a_i \ne a_{i \pm 1}$.

Besides usual multiple polylogarithms we also employ so called cyclotomic polylogarithms
\begin{equation}
\label{MPL_Def_cyclotomic}
G(f_m^l,...,a_n;x)=\int\limits_0^x \frac{x'^lG(a_2,...,a_n;x')}{\Phi_m(x')}dx', \qquad n>0,
\end{equation}
where $\Phi_m(x')$ is a cyclotomic polynomial defined as
\begin{equation}
\Phi_m(x)=\prod_{1 \leqslant k \leqslant m \atop \gcd(k, m) = 1} \left(x - e^{2 \pi i \frac{k}{m}}\right).
\end{equation}
These functions were already extensively studies in the literature and we refer the interested reader to \cite{Ablinger:2011te,Ablinger:2013eba,Kniehl:2018tnp}.
The use of cyclotomic polylogarithm will allow us to write down expansion coefficients of considered hypergeometric functions in much more compact form.
Of course, all cyclotomic polylogarithms can be rewritten in terms of ordinary multiple polylogarithms. For example, we have
\begin{align}
G(\dots,f_4^1,\dots;x) & = \frac{1}{2}\left( G(\dots,i,\dots;x)+G(\dots,-i,\dots;x)\right),
\\
G(\dots,f_4^0,\dots;x) & = \frac{1}{2i}\left(G(\dots,i,\dots;x)-G(\dots,-i,\dots;x)\right).
\end{align}

\section{Differential equation method and  reduction to $\ep$-form}
\label{appendix::DE-epform}

The reduction of Fuchsian differential systems to $\ep$-form is the main instrument used throughout this paper. For this reason let us provide a brief account of its main ideas. Consider matrix differential system
\begin{equation}
\label{eq:DEGeneral}
\frac{d \J}{dx} = \mathcal{\M}(x,\ep) \J
\end{equation}
where $\J$ is a vector of $x$-dependent functions and $\M$ is a matrix rational both in $x$ and $\ep$ variables. In general, the matrix $\M$ may also depend on other parameters. However, it is important that eigenvalues of the matrix residues at singular points over $x$ depend only on $\ep$. The transformation of  vector $\J$ with an invertible matrix $\T$
\begin{equation}
\label{eq:TGeneral}
\J = \T(x,\ep) \Jt
\end{equation}
produces a new differential system
\begin{equation}
\label{eq:DEGeneral2}
\frac{d \Jt}{dx} = \Mt\Jt =\left[\T^{-1}\M \T - \T^{-1}\frac{d}{dx}\T\right]\Jt.
\end{equation}

It was first noticed in \cite{epform1} that sometimes, with the help of transformation \eqref{eq:TGeneral} it is possible to reduce the differential system to a particularly useful form
\begin{equation}
\label{eq:Meneral2}
\Mt(x,\ep) = \ep \sum\limits_r \frac{\Mb_r}{x - x_r} = \ep \Mb(x).
\end{equation}
The latter is known as $\ep$-form. One of the most commonly used algorithms for reducing differential  system to $\ep$-form is the Lee algorithm \cite{epform2}. We also have the criterion for the existence of $\ep$-form in the language of vector bundles over the Riemann sphere \cite{epform-criterium}.
The main advantage of this form is that the perturbative in $\ep$ solution of differential system in this form becomes straightforward.

Lee algorithm for the reduction of differential system to $\ep$-form consists of two main parts.\footnote{There are several mathematical packages implementing Lee algorithm in different programming languages \cite{LeeLibra,Prausa:2017ltv,Gituliar:2017vzm}} The first part of the algorithm brings the differential system to Fuchsian form.\footnote{This means that the Poincaré rank of all singularities of the differential system, including those at infinity, is equal to zero.} and the second part  normalizes the eigenvalues of the matrix residues at singular points. Since we are working with a fixed class of functions, we can initially choose their basis such that corresponding differential system is automatically in Fuchsian form. Therefore, to further reduce the differential system to $\ep$-form we only need the second part of the algorithm, which we will now schematically describe

Suppose we have a differential system with a $\M$-matrix in the Fuchsian form:
\begin{equation}
\label{eq:MFucsian}
\M(x,\ep) = \sum\limits_r \frac{\M_r(\ep)}{x - x_r}.
\end{equation}
To normalize eigenvalues of matrices $\M_r$ at singular points $x_r$ we need a transformation matrix $\T$ that will change them in a controlled manner. As was shown in \cite{epform2} such transformation is provided by the so called balance transformation 
\begin{equation}
\label{eq:balance}
\mathcal{B}(\mathbb{P},x_1,x_2;x) = \mathbb{I} - \mathbb{P} + \frac{x-x_2}{x-x_1}\mathbb{P}\,
\end{equation}
where $\mathbb{P}$ is the projector build from eigenvectors of $\M_1$ and $\M_2^{\top}$ matrices
\begin{equation}
\label{eq:Pforbalance}
\mathbb{P} = \frac{\uvec \wvec^{\intercal}}{\wvec^{\intercal} \uvec}\, ,\qquad \M_1 \uvec =\lambda_1 \uvec\, ,\qquad \wvec^{\intercal}\M_2 = \lambda_2 \wvec^{\intercal}
\end{equation}
This transformation shifts eigenvalues $\lambda_1\to \lambda_1  + 1$ and $\lambda_2\to \lambda_2 - 1$.  So, provided all eigenvalues of the original matrix residues have the form $n+m\ep,~ n \in \mathbb{Z}$ one may build a sequence of balance transformations to make all matrix eigenvalues in the transformed differential system proportional to $\ep$.  After all eigenvalues of the matrix residues became proportional to $\ep$ we need to find extra $x$-independent transformation to explicitly factor out $\ep$-dependence.  
If the transformation matrix $\T$ is $x$-independent then the term $\T^{-1}(x,\ep)\frac{d} {dx}\T(x,\ep)$ in \eqref{eq:DEGeneral2}  disappears and only $\T^{-1}(\ep)\M(\ep)\T(\ep)$ term remains and we may write down the following relation  
\begin{equation}
\label{eq:finalLinearSys}
\frac{\Mt(\ep)}{\ep}\T(\ep,\mu) = \T(\ep,\mu)\frac{\Mt(\mu)}{ \mu}
\end{equation}
where $\T(\ep,\mu) = \T(\ep)\T^{-1}(\mu)$. This linear system for the elements of 
$\T(\ep,\mu)$ matrix is easily solved for a generic $\mu$ and we obtain desired factorization.

The described procedure works if eigenvalues of matrix residues at singular points at $\ep = 0$ were originally integers. If they are not, one may try to find a suitable variable change to make them integer in the transformed differential system.

\section{Diogenes package}
\label{section::Diogenes}
The Diogenes package can be freely downloaded from the bitbucket repository \url{https://bitbucket.org/BezuglovMaxim/diogenes-package/src/master/}. The entire package consists from one file DIOGENES.wl and provided path is set correctly is loaded with the command 
\begin{lstlisting}[language=Mathematica]
<< DIOGENES`
\end{lstlisting}
The main function of the package which expands hypergeometric functions in terms of multiple polylogarithms is \texttt{ExpandHypergeometry}. This function takes three arguments. The first is a hypergeometric function or a combination of them that needs to be expanded. The second is the parameter $\ep$ with respect to which the expansion should be performed and the third one is the expansion order in $\ep$. For example
\begin{mmaCell}[moredefined={ExpandHypergeometry}]{Input}
ExpandHypergeometry[AppellF1[\mmaFrac{1}{2},\(\varepsilon\),\(\varepsilon\),\mmaFrac{1}{2} + \(\varepsilon\),x,y],\(\varepsilon\),1]
\end{mmaCell}
\begin{mmaCell}{Output}
1 + (-G[\{-1\},\mmaSqrt{x}] - G[\{-1\},\mmaSqrt{y}] - G[\{1\},\mmaSqrt{x}] - G[\{1\},\mmaSqrt{y}]) \(\varepsilon\) + \mmaSup{O[\(\varepsilon\)]}{2}
\end{mmaCell}
For those functions that are not included by default in Wolfram Mathematica, we introduce our own notations:  \texttt{AppellF2}, \texttt{AppellF3}, \texttt{AppellF4}, \texttt{LauricellaFA}, \texttt{LauricellaFB} and \texttt{LauricellaFD}.  The arguments of these functions are the same as in their definitions. For Lauricella functions numbered indices and variables are collected into lists.  For example
\begin{mmaCell}[moredefined={ExpandHypergeometry}]{Input}
ExpandHypergeometry[LauricellaFD[\mmaFrac{1}{2} - \(\varepsilon\),\{1,\(\varepsilon\),\(\varepsilon\)\},1 + 2 \(\varepsilon\),\{x,y,z\}],\(\varepsilon\),1]
\end{mmaCell}
\begin{mmaCell}{Output}
\mmaFrac{1}{\mmaSqrt{1-x}} + \(\varepsilon\) (-\mmaFrac{2 G[\{-1\},\mmaSqrt{1-y}]}{\mmaSqrt{1-x}} - \mmaFrac{2 G[\{-1\},\mmaSqrt{1-z}]}{\mmaSqrt{1-x}} + \mmaFrac{6 G[\{0\},\mmaSqrt{1-x}]}{\mmaSqrt{1-x}}
	
+\mmaFrac{2 G[\{-\mmaSqrt{1-y}\},1]}{\mmaSqrt{1-x}} - \mmaFrac{2 G[\{-\mmaSqrt{1-y}\},\mmaSqrt{1-x}]}{\mmaSqrt{1-x}} + \mmaFrac{2 G[\{-\mmaSqrt{1-z}\},1]}{\mmaSqrt{1-x}}
	
-\mmaFrac{2 G[\{-\mmaSqrt{1-z}\},\mmaSqrt{1-x}]}{\mmaSqrt{1-x}} + \mmaFrac{Log[16]}{\mmaSqrt{1-x}}) + \mmaSup{O[\(\varepsilon\)]}{2}
\end{mmaCell}

\begin{mmaCell}[moredefined={ExpandLauricella, \
		LauricellaFAHold}]{Input}
ExpandLauricella[LauricellaFA[\mmaUnd{\(\pmb{\varepsilon}\)},\{\mmaUnd{\(\pmb{\varepsilon}\)},\mmaUnd{\(\pmb{\varepsilon}\)},2\mmaUnd{\(\pmb{\varepsilon}\)},1\},\{1+\mmaUnd{\(\pmb{\varepsilon}\)},1+\mmaUnd{\(\pmb{\varepsilon}\)},1+\mmaUnd{\(\pmb{\varepsilon}\)},1+\mmaUnd{\(\pmb{\varepsilon}\)}\},\{x,y,z,t\}],\mmaUnd{\(\pmb{\varepsilon}\)},2]
\end{mmaCell}
\begin{mmaCell}{Output}
1-\(\varepsilon\) G[\{1\},t]+\mmaSup{\(\varepsilon\)}{2}(G[\{0,1\},t]-G[\{0,1-t\},x]-G[\{0,1-t\},y]-2 G[\{0,1-t\},z])
\end{mmaCell}
There are also special functions \texttt{ExpandPFQ}, \texttt{ExpandAppell} and \texttt{ExpandLauricella}. The first argument of these functions is the corresponding hypergeometric function and the remaining two arguments are the same as those for  \texttt{ExpandHypergeometry}. They also come with more options that make sense for a separate hypergeometric functions. For example, we can get an expansion for the entire function basis

\begin{mmaCell}[moredefined={ExpandAppell, AppellF1Hold}]{Input}
ExpandAppell[AppellF1[\mmaFrac{1}{2},\mmaUnd{\(\pmb{\varepsilon}\)},\mmaUnd{\(\pmb{\varepsilon}\)},\mmaFrac{1}{2}+\mmaUnd{\(\pmb{\varepsilon}\)},x,y],\mmaUnd{\(\pmb{\varepsilon}\)},1,ShowWholeBasis\(\pmb{\to}\)True]
\end{mmaCell}
\begin{mmaCell}{Output}
\{1+\(\varepsilon\) (-G[\{-1\},\mmaSqrt{x}]-G[\{-1\},\mmaSqrt{y}]-G[\{1\},\mmaSqrt{x}]-G[\{1\},\mmaSqrt{y}]),-\mmaFrac{x\(\varepsilon\)}{-1+x},-\mmaFrac{y\(\varepsilon\)}{-1+y}\}
\end{mmaCell}
The correctness of the calculation of the elements of the basis can be checked with the $\theta$ operator, which computes a partial derivative $\theta[f,x] = x \partial f / \partial x$
\begin{mmaCell}[moredefined={G}]{Input}
Simplify[\mmaDef{\(\pmb{\theta}\)}[1+\mmaUnd{\(\pmb{\varepsilon}\)} (-G[\{-1\},\mmaSqrt{x}]-G[\{-1\},\mmaSqrt{y}]-G[\{1\},\mmaSqrt{x}]-G[\{1\},\mmaSqrt{y}]),x]]
\end{mmaCell}
\begin{mmaCell}{Output}
\mmaFrac{x\(\varepsilon\)}{1-x}
\end{mmaCell}
Note that the $\theta$ operator can take derivatives both with respect to the function argument and with respect to its indices. Reduction formulas for Appell functions from  section \ref{section::ReductionAppell} are applied automatically 
\begin{mmaCell}[moredefined={ExpandHypergeometry}]{Input}
AppellF3[\mmaFrac{3}{2} + \(\varepsilon\),-1 - \(\varepsilon\),2 + 3\(\varepsilon\),-\mmaFrac{3}{2} - 3\(\varepsilon\),\mmaFrac{1}{2},x,y]
\end{mmaCell}
\begin{mmaCell}{Output}
\mmaSup{(1 - y)}{3 + 4\(\varepsilon\)} Hypergeometric2F1[\mmaFrac{3}{2} + \(\varepsilon\),2 + 3 \(\varepsilon\),\mmaFrac{1}{2},x + y -xy]
\end{mmaCell}
The cyclotomic kernels  $f_m^l$ are defined as \texttt{f[m,l]}
\begin{mmaCell}[moredefined={ExpandHypergeometry}]{Input}
ExpandHypergeometry[HypergeometricPFQ[\{1,\mmaFrac{1-\mmaUnd{\(\pmb{\varepsilon}\)}}{3}\},\{\mmaFrac{1+\mmaUnd{\(\pmb{\varepsilon}\)}}{3}\},z],\mmaUnd{\(\pmb{\varepsilon}\)},1]
	
"Case C, the solution will be presented in the variable:"  z = \mmaSup{z2[3,z]}{3}
\end{mmaCell}
\begin{mmaCell}{Output}
\mmaFrac{1}{1-\mmaSup{z2[3,z]}{3}}-\mmaFrac{2 \(\varepsilon\)}{3 (-1+\mmaSup{z2[3,z]}{3})} (G[\{1\},z2[3,z]] \mmaSup{z2[3,z]}{2}
	
-2 G[\{f[3,0]\},z2[3,z]] \mmaSup{z2[3,z]}{2}-G[\{f[3,1]\},z2[3,z]] \mmaSup{z2[3,z]}{2})+\mmaSup{O[\(\varepsilon\)]}{2}
\end{mmaCell}
The results for Appell and Lauricella functions the package presents explicitly in terms of radicals, while for generalized hypergeometric functions of one variable compact notation with cyclotomic kernels is used. Cyclotomic polylogarithms can be converted to regular ones using the function  \texttt{ConvertCyclotomicGs}
\begin{mmaCell}[moredefined={ConvertCyclotomicGs, G}]{Input}
ConvertCyclotomicGs[G[\{f[3,0]\},x]]
\end{mmaCell}
\begin{mmaCell}{Output}
\mmaFrac{G[\{-\mmaSup{(-1)}{1/3}\},x]}{1 - 2\mmaSup{(-1)}{1/3}} + \mmaFrac{G[\{\mmaSup{(-1)}{2/3}\},x]}{1 + 2\mmaSup{(-1)}{2/3}}
\end{mmaCell}
This was a brief description of the package functionality. More details can be found in the Mathematica notebook with examples.

\bibliographystyle{hieeetr}
\bibliography{litr}

\end{document}